# Hypericin: Single molecule spectroscopy of an active natural ingredient

Quan Liu,[1,3] Frank Wackenhut,[1*] Otto Hauler,[2] Miriam Scholz,[2] Pierre-Michel Adam,[3] Marc Brecht[1,2] and Alfred J. Meixner[1*]

[1] Institute of Physical and Theoretical Chemistry, Eberhard Karls University Tübingen, Germany

[2] Reutlingen Research Institute, Process Analysis and Technology (PA&T), Reutlingen University, Germany

[3] Laboratoire de Nanotechnologie et d'Instrumentation Optique, Université de Technologie de Troyes, France

**ABSTRACT:** Hypericin can be found in nature in *Hypericum perforatum* (St. John's Wort) and has become subject of intense biochemical research. Studies report of antidepressive, antineoplastic, antitumor and antiviral activity of hypericin. Among the variety of potential applications hypericin can be used as photosensitizer in photodynamic therapy (PDT), where it is brought into cancer cells and produces singlet oxygen upon irradiation with a suitable light source. Therefore, the photophysical properties of hypericin are crucial for a successful application in a medical treatment. Here, we present the first single molecule optical spectroscopy study of hypericin. Its photostability is large enough to obtain single molecule fluorescence, surface enhanced Raman spectra (SERS), fluorescence lifetime, antibunching and blinking dynamics. Embedding hypericin in a PVA matrix changes the blinking dynamics, reduces the fluorescence lifetime and increases the photostability. Single molecule SERS spectra show both the neutral and deprotonated form of hypericin and exhibit sudden spectral changes, which can be associated with a reorientation of the single molecule with respect to the surface.

## Introduction:

Hypericin is a natural active ingredient in *Hypericum perforatum* (St. John's Wort) and has been intensely investigated in biochemical and medical studies during the last decades.[1] Hypericin is reported to be a multifunctional agent in drug and medical application, since it exhibits antidepressive, antineoplastic, antitumor and antiviral activitiy.[1] Many of these applications are based on a phototoxic reaction of hypericin, which can be exploited in photodynamic therapy applied for cancer treatment.[1-5] In photodynamic therapy a photo sensitizer is introduced into the tumor cells and excited with a suitable light source. The photo sensitizer needs to have a high triplet yield to create singlet oxygen, which then destroys the tumor cell. Hypericin is a very promising candidate for such a therapy, since it is the strongest natural photo sensitizer and can additionally be used to visualize tumor cells by its red fluorescence emission.[6] The ensemble properties of hypericin are well documented in several spectroscopic studies.[7-11] However, to the best of our knowledge, single molecule (SM) studies of hypericin are missing. SM spectroscopy can contribute to a better understanding of its photophysical properties, since certain aspects, e.g. influence of the local environment or subpopulations, can be hidden by ensemble averaging.[12-17] Furthermore, SM experiments allow to investigate dynamics caused by changes of the local environment or of the structure or orientation of the single molecule.[18-23] This possibility to observe individual processes and reassemble ensemble signals makes SM experiments a powerful tool to investigate spectroscopic properties, which is crucial for a medical application or advanced microscopy techniques like fluorescence lifetime imaging, PALM or spt-PALM.

In this work, we present the first SM experiments with hypericin and will show that it is possible to study its optical properties, e.g. fluorescence blinking, fluorescence lifetimes and single molecule surface enhanced Raman spectroscopy (SERS), despite its high triplet yield of 0.5-0.7.[7, 24] Furthermore, we will investigate the influence of the embedding matrix on these optical properties.

## Results and Discussion

Hypericin is an anthrachinon derivate with its molecular structure depicted in Fig. 1 (a).

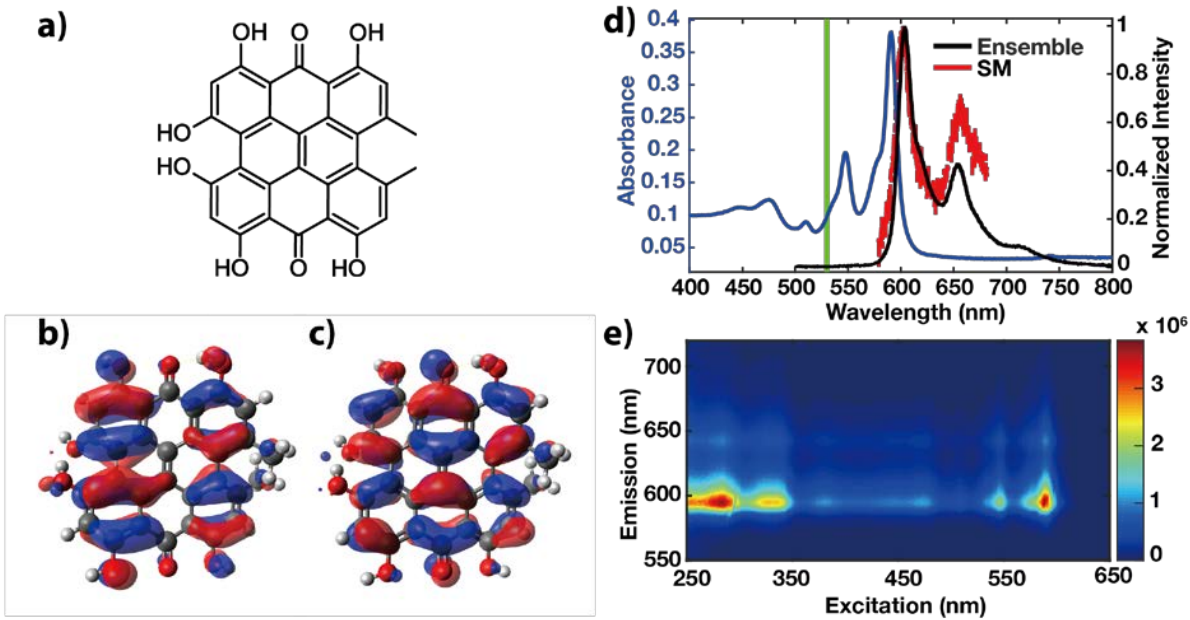

**Figure.1** (a) Chemical structure of hypericin. (b)/(c) HOMO and LUMO of hypericin. (d) Ensemble absorption (blue line) and fluorescence (black line) spectra of hypericin in ethanol as well as an emission spectrum of a single hypericin molecule (red line) on a glass coverslip. The excitation wavelength was 530 nm for both emission spectra and is indicated by the green line. (e) Combined excitation and fluorescence spectrum of hypericin.

Fig 1 (b)/(c) illustrates ab initio calculations of the HOMO and LUMO of hypericin, respectively. All ab initio calculations in this work have been performed with the Gaussian® 09 program.[25] The used method is the B3LYP hybrid exchange correlation functional, since it has been benchmarked for many systems over decades and is therefore highly viable for the simulations of aromatic organic molecules with complex functional groups.[26-29] The used basic set was a split-valence basis (6-311++G*), which has been shown to perform well for describing second-row elements,[30, 31] while being well suited to simulate Raman scattering cross-sections.[32] Initial calculations have been carried out with the UHF/6-31G method to obtain optimized geometries for free hypericin. This geometry has been used as initial guess for a full geometry optimization of the molecule utilizing the beforehand mentioned basis set and method with a tight criterion. These optimized geometries permit the calculation of the vibrational modes of hypericin using an ultrafine grid in the calculation of the integrals, as well as a population analysis yielding the molecular orbitals presented in Fig. 1(b)/(c). The resulting wavenumbers of the vibrational normal modes have been scaled by the factor of 0.967 to address the systematic errors of the harmonic frequency analysis.[33]

The characterization of the optical properties of hypericin on the ensemble level was done by acquiring the absorption (blue line) and the fluorescence (black line) spectrum, which is shown in Fig. 1(d). The absorption spectrum (blue line) can be roughly divided into three different domains, i.e. the absorption to the $S_1$ state between 490 nm and 600 nm (with the electronic transition $S_0 \rightarrow S_1$ at 590.83 nm and its vibronic manifold), the $S_2$ absorption between 400 nm and 490 nm (electronic transition $S_0 \rightarrow S_2$ at 480 nm) and the triplet-triplet absorption at 510 nm.[11, 34] The respective fluorescence spectrum was recorded with a pulsed excitation laser $\lambda_{ex}$=530 nm (Fig.1 d green line), a 532 nm long pass filter (Semrock 532LP Edge Basic) and an integration time of 1s. The ensemble fluorescence spectrum (black line) has mirror symmetry to the absorption spectrum and the first $S_1$ emission peak is redshift by ~13 nm to 603 nm. The red line in Fig. 1(d) displays an exemplary fluorescence spectrum of a single hypericin molecule showing the same emission bands like the ensemble spectrum, but with different intensity ratios. The concentration of hypericin in the SM experiment was adjusted such that the average spatial distance between single hypericin molecules was much larger than the diffraction limited focal spot of the confocal microscope to ensure that only one molecule was within the focal volume at a time. Within the 2D plot in Fig 1(e), a combined excitation and fluorescence spectrum of hypericin is illustrated. It coincides with the previously described absorption and emission transitions.

To study the influence of the local environment on single hypericin molecules two different types of samples were prepared, hypericin directly spin coated on a carefully cleaned glass substrate and hypericin embedded in a 2wt% PVA matrix (for details see material and methods section). Fluorescence intensity images of single hypericin molecules on glass and embedded in PVA are shown in Fig.2 (a)/(b), respectively.

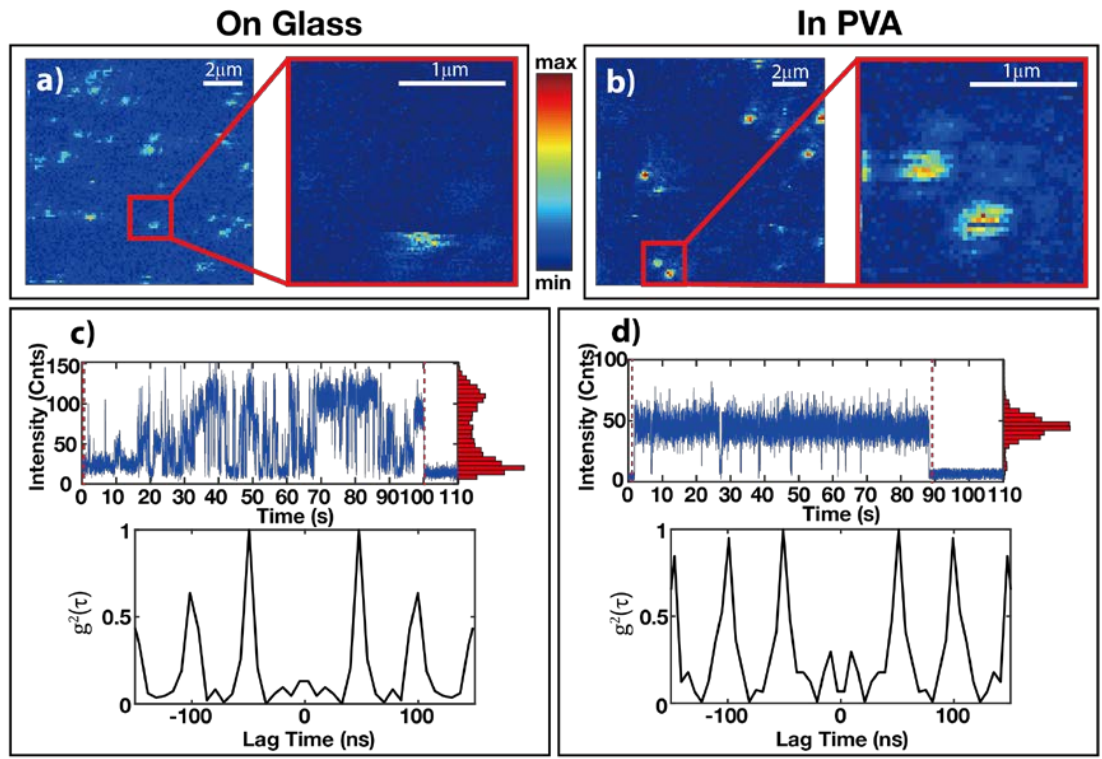

**Figure.2** Fluorescence properties of single hypericin molecules. **(a,b)** Fluorescence images of single hypericin molecules on a glass coverslip and embedded in 2wt% PVA. The inset is a close-up scan and shows the characteristic bleaching and blinking behavior of single molecules. The upper row in **(c,d)** shows two exemplary intensity time traces of single hypericin molecule on glass **(c)** and embedded in PVA **(d)**, the corresponding histograms can be seen on the right side. The lower row shows the second-order intensity correlation function calculated from the respective intensity time traces shown in the upper row.

The fluorescence images were recorded with a raster scanning confocal microscope with pulsed excitation ($\lambda_{ex} = 530\ nm$, 1.4 µW, 20 MHz), a 532 nm long pass filter and an integration time of 5 ms per pixel. The images show isolated Gaussian shaped diffraction limited spots, more details can be observed in the close up views on the right side of Fig.2 (a)/(b). Intensity fluctuations in the image patterns occur since the molecule is in a dark state for a certain amount of time, e.g. the triplet state or in another non-emissive states.[35-38] This blinking is characteristic for a SM and, in addition to the single step photo bleaching, gives an indication that a SM is observed. In order to investigate the blinking dynamic of hypericin, we have recorded intensity time traces of SMs. Two examples, with a binning time of 10ms, are shown in the upper row of Fig.2 (c)/(d). The intensity time traces were recorded until the molecule photo bleached, which was confirmed when no emission signal could be recorded after at least 20 s. The two exemplary intensity time traces show the characteristic on/off behavior of a SM since it can be either in a bright or in a dark state, which was already observed in the images in Fig.2 (a)/(b). Interestingly, hypericin on the glass substrate in Fig.2 (c) does not show a strict single step blinking, which can be caused, e.g. by changes of the molecular geometry such as intramolecular proton transfer or transitions between different tautomers.[18, 19, 24, 39-41] Furthermore, the two exemplary time traces suggest that there are distinct differences between the two environments, which becomes more obvious in the histograms shown on the right side of the corresponding time trace. The SM on the glass substrate is more often in the dark state, causing strong intensity fluctuations, while the SM in the PVA matrix is mainly in the bright state until it spontaneously photobleaches. The second-order photon correlation functions of the time traces shown in Fig.2 (c)/(d) are presented below the corresponding intensity time trace. The absence of the peak at $\tau = 0\ s$ is caused by antibunching and confirms that the observed emission is indeed caused by a SM. To study the survival time and blinking dynamics in more detail the intensity time traces of single hypericin molecules were analyzed statistically. The survival time, i.e. the time before a molecule photo bleaches due to the repeated excitation, of single hypericin molecules is presented in Fig. 3(a)/(b).

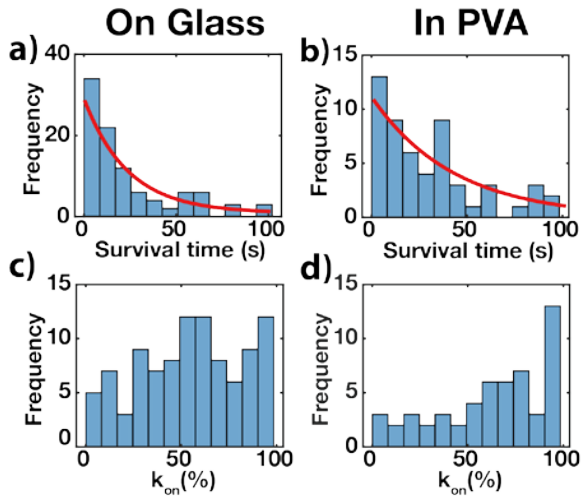
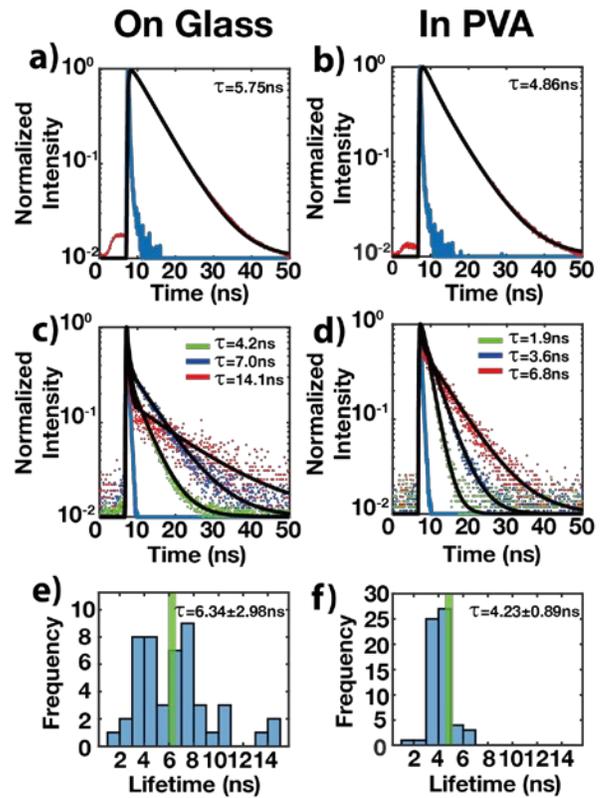

**Figure.3** Statistics of single hypericin molecule blinking behavior. **(a,b)** show the histogram of the survival times of single hypericin molecules on a glass substrate (n=98) **(a)** and in a PVA matrix (n=54) **(b)**. Interestingly, single hypericin molecules show a remarkable photostability and can live for as long as hundred seconds. The red line shows an exponential fit to the histograms and gives a bleaching time of 23.8s on the glass substrate and 46.6s in the PVA matrix. **(c,d)** On-ratio $k_{on}$ histogram of single hypericin molecules on a glass substrate **(c)** and in a PVA matrix **(d)**. On the glass substrate the hypericin molecules show strong blinking, while it is suppressed in the PVA matrix.

Fig. 3(a) illustrates the survival time for 98 hypericin molecules on a glass substrate, while the results shown in Fig. 3(b) are based on 54 hypericin molecules in a PVA matrix. In both cases we find survival times reaching from a few seconds up to 100s. Using an exponential fit function (red line in Fig. 3 (a)/(b)), we determine an average survival time of 23.8s on the glass substrate (Fig. 3 (a)) and 46.6s in the PVA matrix (Fig. 3 (b)). In consequence, due to this increase of the average survival time the portion of long living molecules is larger in PVA. The same intensity time traces were used to investigate the blinking dynamics by determining the on-ratio $k_{on}$, *i.e.* the ratio between the time when the molecule is in the bright $t_{on}$ or in the dark $t_{off}$ state. For this purpose, an on/off threshold was set for each intensity time trace and the molecule is considered to be in the bright state when the fluorescence intensity is above this threshold. This allows to determine $t_{on}$ and $t_{off}$ and consequently $k_{on}$ can be calculated by $k_{on} = (t_{on}/(t_{on} + t_{off}))\times100$. Fig. 3 (c) presents the results for 98 molecules on the glass substrate and $k_{on}$ is almost evenly distributed from close to 0% up to 100%. The blinking dynamic changes when hypericin is embedded in the PVA matrix, which is shown for 54 SMs in Fig. 3(d). Clearly, the amount of single hypericin molecules which hardly show any blinking increases with a significant amount of molecules having $k_{on}$ close to 100%.

Such a change in the blinking dynamic, observed in Fig. 3(c)/(d), is often accompanied by a difference in the fluorescence lifetime. Thus, we performed SM Time Correlated Single Photon Counting (TCSPC) for hypericin on glass and embedded in a PVA matrix. The results are shown in Fig. 4.

**Figure.4** Fluorescence lifetime of hypericin. **(a,b)** show ensemble time correlated single photon counting (TCSPC) traces of hypericin on a glass substrate **(a)** and in a PVA matrix **(b)**. The experimental decay curves are shown in red together with the instrument response function (blue line). The excited state lifetimes determined from the exponential fit (black line) are 5.75ns and 4.86ns, respectively. **(c,d)** display TCSPC traces and fitting results of three single hypericin molecule on glass **(c)** and in a PVA matrix **(d)**. The corresponding lifetime histograms are presented in **(e,f)**. The green line indicates the ensemble lifetime and the respective standard deviations are $6.34\pm2.98$ ns (n=48) for the glass substrate and $4.23\pm0.89$ ns (n=61) in the PVA matrix.

Fig. 4 (a)/(b) shows fluorescence decay curves (red dots) of an ensemble of hypericin on a glass substrate (a) and embedded in a PVA matrix (b). Average fluorescence lifetimes were obtained by fitting the decay curves with a double exponential function (black line), which is convoluted with the instrument response function (IRF, blue line). The average fluorescence lifetime on the glass substrate is 5.75ns and 4.86ns in the PVA matrix. These values are consistent with fluorescence lifetimes reported for ensemble hypericin in different solutions.[42] Exemplary fluorescence decay curves of three SMs and their respective exponential fitting functions (black lines) are displayed in Fig. 4 (c)/(d). On the SM level, the fluorescence lifetime varies considerably. The fluorescence lifetimes are ranging from 1.65 ns to 14.65 ns on the glass substrate and from 1.91 ns to 6.95 ns in the PVA matrix. Corresponding histograms for n=48 molecules on the glass substrate and n=61 in the PVA matrix are displayed in Fig. 4 (e)/(f), respectively. An average fluorescence lifetime of $\tau = (6.34 \pm 2.98)\ ns$ on the glass substrate and of $\tau = (4.23 \pm 0.89)\ ns$ in the PVA matrix can be determined from these histograms. This coincides well with the ensemble fluorescence lifetime in Fig. 4 (a)/(b), which is indicated by the green line in Fig. 4(e)/(f). However, there is an obvious difference in the standard deviation of the SM fluorescence lifetimes for the two sample types, since the standard deviation reduces from 2.98ns on the glass substrate to 0.89ns in the PVA matrix. The shorter fluorescence lifetimes in the PVA matrix lead to a larger radiative decay rate, which can also be observed by the higher brightness in the fluorescence images in Fig. 2 (a)/(b). Additionally, a shorter fluorescence lifetime reduces the chance of a transition to the triplet state and blinking and bleaching is also reduced, which can be clearly observed in Fig. 3. Consequently, the results in Fig. 2-4 show that the PVA host matrix has a stabilizing effect and that the matrix alters the spectroscopic properties of hypericin remarkably, which has also been shown for e.g. Photosystem I.[43] Altogether, hypericin is highly sensitive to the local environment making it an ideal candidate for applications such as fluorescence lifetime imaging.

Hypericin was additionally characterized with Surface Enhanced Raman Scattering (SERS). For this purpose, a silver island film functions as SERS substrate, which is depicted in the electron microscopy image shown on the left side of Fig. 5(a).

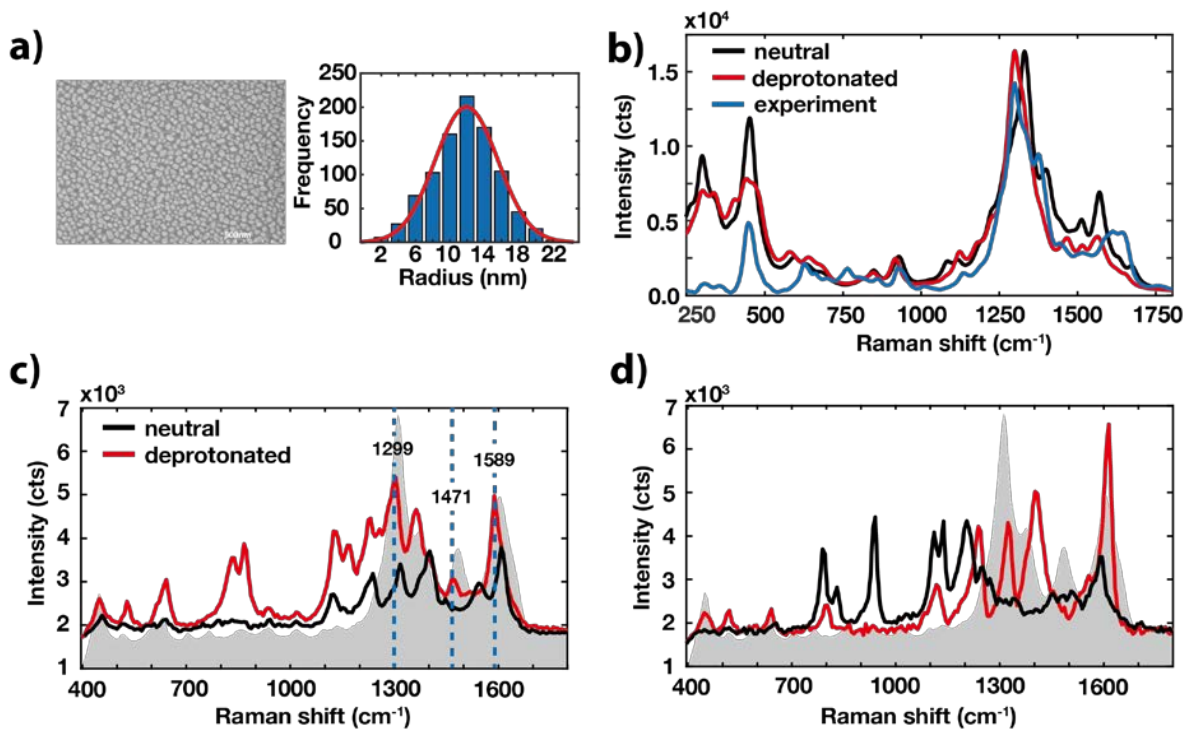

**Figure.5** Surface Enhanced Raman Scattering (SERS) spectra of hypericin. **(a)** displays an electron microscopy image of the SERS substrate consisting of a silver island film on a glass substrate. The corresponding particle size distribution of the substrate is R = $(11.93 \pm 3.66)$ nm **(b)** Ensemble SERS spectra acquired with an excitation wavelength of $\lambda_{ex}$=633nm (blue line) and DFT simulation of the neutral (black line) and deprotonated (red line) form of hypericin. Several pronounced peaks are resolved in 1200~1700 cm$^{-1}$ region. **(c)** SM SERS spectra excited at $\lambda_{ex}$=530nm. The black and red line show SM SERS spectra of the neutral and deprotonated form of hypericin, respectively. The blue dashed lines mark prominent peaks of deprotonated hypericin. The grey area displays the ensemble resonance Raman spectrum. **(d)** SERS spectrum of the same neutral hypericin SM acquired at different times. The intensity differences between the red and black spectrum indicate, that a spontaneous reorientation of the SM with respect to the nanoparticle surface did occur. The grey area displays the ensemble resonance Raman spectrum.

The electron micrographs were analyzed with with the software "ImageJ" and the size distribution of the nanoparticle film in Fig. 5 (a) is $(11.93 \pm 3.66)\ nm$, as determined from the histogram shown on the right side. An extinction spectrum of the SERS substrate is presented in the supporting information Fig. S1 and shows a particle plasmon resonance at 482 nm. 5μL of a 1μM hypericin solution was dropped on such a substrate to acquire ensemble SERS spectra. For the ensemble experiments we used a 633nm excitation laser, which is not resonant with an electronic transition of hypericin. The results are shown in black in Fig. 5(b). The SERS spectra are analyzed with a self-written Matlab script and the corresponding peak positions are listed in Table 1 of the supporting information. They are consistent with values reported in literature for hypericin.[8, 10, 11, 44] Additionally, DFT calculations were performed and respective simulated spectra for the neutral and deprotonated form of hypericin are shown in black and red in Fig. 5 (b), respectively. The first deprotonation of hypericin occurs at the *bay* hydroxyl groups, which have a low pKa of 1.5.[10, 11] Raman peak positions agree well with the simulated and the experimental ensemble spectrum. Comparing the simulated spectra with the experimental ensemble spectrum suggests that both the neutral and deprotonated species of hypericin are present in the ensemble experiment. The pronounced peaks in the SERS spectrum at 1297 cm$^{-1}$ and at 1471 cm$^{-1}$ are assigned to the deprotonated form of hypericin. On the other hand, the shoulder at ~1333 cm$^{-1}$ and the peak at 1380 cm$^{-1}$ in the ensemble SERS spectrum originate from the neutral species. SERS experiments on the SM level allow us to differentiate between the neutral and deprotonated form of hypericin within the same specimen. SM SERS spectra of hypericin were acquired by using the same SERS substrates, but the concentration of the hypericin solution was reduced to 0.1 nM. SERS images recorded from pristine substrates and from substrates doped with 1.0nM and 0.1nM hypericin solution are presented in the supporting information Fig. S2 (a)-(c), respectively. The SM SERS spectra were acquired from the intense hot spots shown in Fig. S2 (b)-(c). The SM SERS spectra were measured with a 530nm laser, which is in resonance with the $S_0 \rightarrow S_1$ electronic transition of hypericin enabling resonance Raman scattering. For comparison the ensemble resonance Raman spectrum is displayed by the grey area in Fig. 5(c)/(d) and the strongest enhancement due to the resonance effect occurs for Raman peaks between 1400 cm$^{-1}$ and 1600 cm$^{-1}$. Two exemplary SM SERS spectra are shown in Fig. 5(c). In general, the same SERS peaks as in the ensemble spectrum can be observed for SMs, but the full width at half maximum of the peaks is significantly reduced since there is no ensemble averaging. The SERS spectrum shown in Fig. 5(c) in red exhibits strong peaks at 1589 cm$^{-1}$, 1471 cm$^{-1}$ and 1299 cm$^{-1}$ (marked by the dashed lines). These peaks are also present in the ensemble spectrum in Fig. 5(b) and have been assigned to the deprotonated form of hypericin.[10] However, the relative intensity of these peaks is strongly reduced in the spectrum of neutral hypericin (black, Fig.5(c)) compared to the deprotonated hypericin molecule (red, Fig.5(c)) and the peak positions are shifted to 1610 cm$^{-1}$, 1451 cm$^{-1}$ and 1316 cm$^{-1}$. These changes of the SERS spectrum can be attributed to the deprotonation of hypericin.[10] Therefore, SM SERS experiments allow to directly observe different hypericin subpopulations and show that both the neutral and deprotonated species of hypericin are adsorbed on the SERS substrate. Furthermore, comparing these SM spectra with the ensemble it is evident, that the ensemble is a mixture of neutral and deprotonated hypericin. The spectral peak positions of these SERS experiments are summarized in Table 1 in the supporting information for the ensemble, the deprotonated and the neutral SM together with the corresponding simulations. The SERS spectra presented in Fig. 5(d) are obtained from the same SM at different times and strong intensity variations can be observed between the spectra shown in red and black. The strongest intensity differences occur between 1400-1600cm$^{-1}$ and the peaks in this spectral region have an in plane character, which makes them very sensitive to the orientation of the SM on the nanoparticle surface.[8, 10] The near field of the silver nanoparticle is oriented perpendicular to its surface and Raman modes with a Raman tensor component parallel to the field are enhanced.[45] This makes it possible to determine the orientation of the SM relative to the surface. The changes in the SERS intensities indicate that the same SM is first oriented almost parallel (black line in Fig. 5(d)) and reorients perpendicularly or almost perpendicularly to the nanoparticle surface (red line in Fig. 5(d)). Such a perpendicular orientation of hypericin on the silver surface is also proposed in ensemble SERS studies.[8, 9] Altogether, SM SERS experiments show that both the neutral and deprotonated form of hypericin are adsorbed on the silver surface and indicate that hypericin can reorient on the silver surface.

**Summary**

We report the first single molecule study of hypericin and show that it is, despite its high triplet yield, excellently suited for fluorescence spectroscopy and SERS on the SM level. We have shown that the photostability of hypericin is large enough to acquire SM fluorescence and SERS spectra. The SM SERS experiments show that both the neutral and deprotonated form of hypericin is adsorbed on the silver nanoparticle and that it can change its orientation on the surface. Furthermore, hypericin seems to be ideal to examine variations in the local environment, since its photophysical properties, e.g. blinking dynamic, fluorescence lifetime and SERS spectra, strongly depend on the local environment of the SM. Especially, the large variation in the fluorescence lifetime makes it ideal for imaging applications like fluorescence lifetime imaging. Furthermore, these new insights on the optical properties of single hypericin molecules can be utilized to improve medical applications, e.g. photodynamic diagnosis (PDD) or photodynamic therapy (PDT), and will help to develop new medical therapies based on this natural active ingredient.

**Methods and Materials**

Hypericin and polyvinylalcohol (PVA) were purchased from Sigma Aldrich. Hypericin was dissolved in ethanol (UVsol) and diluted to a concentration of 10$^{-10}$ mol/l for the fluorescence and SERS experiments. PVA was dissolved in triple distilled water and 2μL of hypericin solution was added to a 2wt% PVA solution to obtain the required concentration. All solutions were stocked in the dark and at 4°C to minimize molecule bleaching. Coverslips were cleaned with chromosulfuric acid solution, rinsed with triple distilled water and finally dried in a nitrogen flow.

The SERS substrate was prepared by evaporating (Edwards EB3, 1e-6 mbar, 0.1nm/s) a 4nm thick silver layer on cleaned coverslips, resulting in a silver island film with an average particle size of 11.93±3.66 nm.[46] On this silver island film, a 5 μL droplet of 1μM hypericin solution was directly added to acquire the ensemble SERS spectra shown in Fig. 5(b). A 1e-10 M concentrated hypericin solution was used to acquire the single molecule SERS spectra shown in Fig. 5(c)/(d).

Ensemble fluorescence (Fig.1 (b)) and fluorescence lifetime measurements (Fig.4 (a)/(b)) were performed with a 20 μl spin-coated sample of a 1e-5M hypericin solution with and without PVA on clean coverslips. The same solution was used for the ensemble absorption measurement (Fig.1 (b)), which was recorded with a UV-VIS-NIR spectrophotometer (PerkinElmer Lambda 19).

The 2D plot in Fig. 1(e) was recorded with a Fluorolog®-3 instrument (Horiba Jobin Yvon GmbH) in right-angle detection mode with 1nm spectral steps of the excitation and emission monochromator and a corresponding integration time of 0.1s per spectrum. Monochromator bandwidths were adjusted to 1nm.

For single molecule imaging (Fig.2 (a)/(b)), time trace, and antibunching (Fig.2 (c)/(d)) and lifetime measurements (Fig.4 (c)/(d)), we used 5 µL of 1e-10 M hypericin added directly on a coverslip or 20 µL of 1e-9 M hypericin spin coated (6k rpm, 30s) on a coverslip.

Fluorescence images/spectra and lifetimes of single hypericin molecules were recorded with a home built scanning confocal microscope.[47] A 530nm pulsed laser (1.4µW, 20 MHz) was used to illuminate the sample with a high numerical aperture (NA=1.46) oil objective lens. The fluorescence signal is collected by the same objective lens and sent to an avalanche photodiode, which is connected to a Time-Correlated Single Photon Counting Detector (TCSP, HydraHarp 400). Decay curves were fitted and analyzed with SymPhoTime 64.

## AUTHOR INFORMATION

**Corresponding Author**

* frank.wackenhut@uni-tuebingen.de
* alfred.meixner@uni-tuebingen.de

**Funding Sources**

The authors acknowledge funding of the DFG (ME 1600/13-3) and the Funding of China Scholarship Council (CSC).

## REFERENCES

1. Kubin, A.; Wierrani, F.; Burner, U.; Alth, G.; Grunberger, W., Hypericin - The Facts About a Controversial Agent. *Current Pharmaceutical Design* **2005,** *11* (2), 233-253.
2. Agostinis, P.; Vantieghem, A.; Merlevede, W.; de Witte, P. A. M., Hypericin in cancer treatment: more light on the way. *The International Journal of Biochemistry & Cell Biology* **2002,** *34* (3), 221-241.
3. Davids, L. M.; Kleemann, B.; Kacerovská, D.; Pizinger, K.; Kidson, S. H., Hypericin phototoxicity induces different modes of cell death in melanoma and human skin cells. *Journal of Photochemistry and Photobiology B: Biology* **2008,** *91* (2), 67-76.
4. Vandenbogaerde, A. L.; Kamuhabwa, A.; Delaey, E.; Himpens, B. E.; Merlevede, W. J.; de Witte, P. A., Photocytotoxic effect of pseudohypericin versus hypericin. *Journal of Photochemistry and Photobiology B: Biology* **1998,** *45* (2), 87-94.
5. Ritz, R.; Scheidle, C.; Noell, S.; Roser, F.; Schenk, M.; Dietz, K.; Strauss, W. S., In vitro comparison of hypericin and 5-aminolevulinic acid-derived protoporphyrin IX for photodynamic inactivation of medulloblastoma cells. *PLoS One* **2012,** *7* (12), e51974.
6. Ritz, R.; Müller, M.; Weller, M.; Dietz, K.; Kuci, S.; Roser, F.; Tatagiba, M., Hypericin: a promising fluorescence marker for differentiating between glioblastoma and neurons in vitro. *International journal of oncology* **2005,** *27* (6), 1543-1549.
7. Darmanyan, A. P.; Jenks, W. S.; Eloy, D.; Jardon, P., Quenching of Excited Triplet State Hypericin with Energy Acceptors and Donors and Acceptors of Electrons. *The Journal of Physical Chemistry B* **1999,** *103* (17), 3323-3331.
8. Jancura, D.; Sánchez-cortés, S.; Kocisova, E.; Tinti, A.; Miskovsky, P.; Bertoluzza, A., Surface-enhanced resonance raman spectroscopy of hypericin and emodin on silver colloids: SERRS and NIR FTSERS study. *Biospectroscopy* **1995,** *1* (4), 265-273.
9. Raser, L. N.; Kolaczkowski, S. V.; Cotton, T. M., RESONANCE RAMAN AND SURFACE-ENHANCED RESONANCE RAMAN SPECTROSCOPY OF HYPERICIN. *Photochemistry and Photobiology* **1992,** *56* (2), 157-162.
10. Lajos, G.; Jancura, D.; Miskovsky, P.; García-Ramos, J. V.; Sanchez-Cortes, S., Surface-Enhanced Fluorescence and Raman Scattering Study of Antitumoral Drug Hypericin: An Effect of Aggregation and Self-Spacing Depending on pH. *The Journal of Physical Chemistry C* **2008,** *112* (33), 12974-12980.
11. Wynn, J. L.; Cotton, T. M., Spectroscopic properties of hypericin in solution and at surfaces. *The Journal of Physical Chemistry* **1995,** *99* (12), 4317-4323.
12. Dickson, R. M.; Cubitt, A. B.; Tsien, R. Y.; Moerner, W. E., On/off blinking and switching behaviour of single molecules of green fluorescent protein. *Nature* **1997,** *388* (6640), 355-358.
13. Brecht, M.; Studier, H.; Radics, V.; Nieder, J. B.; Bittl, R., Spectral Diffusion Induced by Proton Dynamics in Pigment–Protein Complexes. *Journal of the American Chemical Society* **2008,** *130* (51), 17487-17493.
14. Nieder, J. B.; Brecht, M.; Bittl, R., Dynamic intracomplex heterogeneity of phytochrome. *Journal of the American Chemical Society* **2008,** *131* (1), 69-71.
15. Nieder, J. B.; Bittl, R.; Brecht, M., Fluorescence Studies into the Effect of Plasmonic Interactions on Protein Function. *Angewandte Chemie International Edition* **2010,** *49* (52), 10217-10220.
16. Basché, T.; Moerner, W.; Orrit, M.; Wild, U., *Single-molecule optical detection, imaging and spectroscopy*. John Wiley & Sons: 2008.
17. Lu, H. P.; Xie, X. S., Single-molecule spectral fluctuations at room temperature. *Nature* **1997,** *385* (6612), 143-146.
18. Stracke, F.; Blum, C.; Becker, S.; Müllen, K.; Meixner, A. J., Intrinsic conformer jumps observed by single molecule spectroscopy in real time. *Chemical Physics Letters* **2000,** *325* (1), 196-202.
19. Blum, C.; Stracke, F.; Becker, S.; Müllen, K.; Meixner, A. J., Discrimination and Interpretation of Spectral Phenomena by Room-Temperature Single-Molecule Spectroscopy. *The Journal of Physical Chemistry A* **2001,** *105* (29), 6983-6990.
20. Piatkowski, L.; Schanbacher, C.; Wackenhut, F.; Jamrozik, A.; Meixner, A. J.; Waluk, J., Nature of Large Temporal Fluctuations of Hydrogen Transfer Rates in Single Molecules. *The Journal of Physical Chemistry Letters* **2018,** *9* (6), 1211-1215.
21. Vosgröne, T.; Meixner, A. J., Surface- and Resonance-Enhanced Micro-Raman Spectroscopy of Xanthene Dyes: From the Ensemble to Single Molecules. *ChemPhysChem* **2005,** *6* (1), 154-163.
22. Schlücker, S., Surface-Enhanced Raman Spectroscopy: Concepts and Chemical Applications. *Angewandte Chemie International Edition* **2014,** *53* (19), 4756-4795.
23. Chizhik, A. M.; Jäger, R.; Chizhik, A. I.; Bär, S.; Mack, H.-G.; Sackrow, M.; Stanciu, C.; Lyubimtsev, A.; Hanack, M.; Meixner, A. J., Optical imaging of excited-state tautomerization in single molecules. *Physical Chemistry Chemical Physics* **2011,** *13* (5), 1722-1733.
24. Falk, H., From the Photosensitizer Hypericin to the Photoreceptor Stentorin— The Chemistry of Phenanthroperylene Quinones. *Angewandte Chemie International Edition* **1999,** *38* (21), 3116-3136.
25. Frisch, M.; Trucks, G.; Schlegel, H. B.; Scuseria, G.; Robb, M.; Cheeseman, J.; Scalmani, G.; Barone, V.; Mennucci, B.; Petersson, G., Gaussian 09, revision a. 02, gaussian. *Inc., Wallingford, CT* **2009,** *200*, 28.
26. Sousa, S. F.; Fernandes, P. A.; Ramos, M. J., General Performance of Density Functionals. *The Journal of Physical Chemistry A* **2007,** *111* (42), 10439-10452.
27. Krishnakumar, V.; Keresztury, G.; Sundius, T.; Ramasamy, R., Simulation of IR and Raman spectra based on scaled DFT force fields: a case study of 2-(methylthio)benzonitrile, with emphasis on band assignment. *Journal of Molecular Structure* **2004,** *702* (1), 9-21.
28. Negri, F.; Castiglioni, C.; Tommasini, M.; Zerbi, G., A Computational Study of the Raman Spectra of Large Polycyclic Aromatic Hydrocarbons: Toward Molecularly Defined Subunits of Graphite. *The Journal of Physical Chemistry A* **2002,** *106* (14), 3306-3317.
29. Polavarapu, P. L., Ab initio vibrational Raman and Raman optical activity spectra. *The Journal of Physical Chemistry* **1990,** *94* (21), 8106-8112.
30. McLean, A. D.; Chandler, G. S., Contracted Gaussian basis sets for molecular calculations. I. Second row atoms, Z=11–18. *The Journal of Chemical Physics* **1980,** *72* (10), 5639-5648.
31. Krishnan, R.; Binkley, J. S.; Seeger, R.; Pople, J. A., Self‐consistent molecular orbital methods. XX. A basis set for correlated wave functions. *The Journal of Chemical Physics* **1980,** *72* (1), 650-654.

32. Cheeseman, J. R.; Frisch, M. J., Basis Set Dependence of Vibrational Raman and Raman Optical Activity Intensities. *Journal of Chemical Theory and Computation* **2011,** *7* (10), 3323-3334.
33. Johnson III, R. D., *NIST Computational Chemistry Comparison and Benchmark Database*, Accessed on: 18. April 2018.
34. Arabei, S. M.; Galaup, J. P.; Jardon, P., Analysis of the site selected fluorescence and the phosphorescence spectrum of hypericin in ethanol. *Chemical Physics Letters* **1997,** *270* (1), 31-36.
35. Panzer, O.; Göhde, W.; Fischer, U. C.; Fuchs, H.; Müllen, K., Influence of Oxygen on Single Molecule Blinking. *Advanced Materials* **1998,** *10* (17), 1469-1472.
36. Zondervan, R.; Kulzer, F.; Orlinskii, S. B.; Orrit, M., Photoblinking of Rhodamine 6G in Poly(vinyl alcohol): Radical Dark State Formed through the Triplet. *The Journal of Physical Chemistry A* **2003,** *107* (35), 6770-6776.
37. Clifford, J. N.; Bell, T. D. M.; Tinnefeld, P.; Heilemann, M.; Melnikov, S. M.; Hotta, J.-i.; Sliwa, M.; Dedecker, P.; Sauer, M.; Hofkens, J.; Yeow, E. K. L., Fluorescence of Single Molecules in Polymer Films: Sensitivity of Blinking to Local Environment. *The Journal of Physical Chemistry B* **2007,** *111* (25), 6987-6991.
38. Stracke, F.; Blum, C.; Becker, S.; Müllen, K.; Meixner, A. J., Correlation of Emission Intensity and Spectral Diffusion in Room Temperature Single-Molecule Spectroscopy. *ChemPhysChem* **2005,** *6* (7), 1242-1246.
39. English, D. S.; Das, K.; Ashby, K. D.; Park, J.; Petrich, J. W.; Castner, E. W., Confirmation of Excited-State Proton Transfer and Ground-State Heterogeneity in Hypericin by Fluorescence Upconversion. *Journal of the American Chemical Society* **1997,** *119* (48), 11585-11590.
40. Piwoński, H.; Hartschuh, A.; Urbańska, N.; Pietraszkiewicz, M.; Sepioł, J.; Meixner, A. J.; Waluk, J., Polarized Spectroscopy Studies of Single Molecules of Porphycenes: Tautomerism and Orientation. *The Journal of Physical Chemistry C* **2009,** *113* (27), 11514-11519.
41. Piwoński, H.; Stupperich, C.; Hartschuh, A.; Sepioł, J.; Meixner, A.; Waluk, J., Imaging of Tautomerism in a Single Molecule. *Journal of the American Chemical Society* **2005,** *127* (15), 5302-5303.
42. Yamazaki, T.; Ohta, N.; Yamazaki, I.; Song, P. S., Excited-state properties of hypericin: electronic spectra and fluorescence decay kinetics. *The Journal of Physical Chemistry* **1993,** *97* (30), 7870-7875.
43. Hussels, M.; Brecht, M., Effect of Glycerol and PVA on the Conformation of Photosystem I. *Biochemistry* **2011,** *50* (18), 3628-3637.
44. Miškovský, P.; Jancura, D.; Sánchez-Cortés, S.; Kočišová, E.; Chinsky, L., Antiretrovirally Active Drug Hypericin Binds the IIA Subdomain of Human Serum Albumin: Resonance Raman and Surface-Enhanced Raman Spectroscopy Study. *Journal of the American Chemical Society* **1998,** *120* (25), 6374-6379.
45. Creighton, J. A., Surface raman electromagnetic enhancement factors for molecules at the surface of small isolated metal spheres: The determination of adsorbate orientation from sers relative intensities. *Surface Science* **1983,** *124* (1), 209-219.
46. Semin, D. J.; Lo, A.; Roark, S. E.; Skodje, R. T.; Rowlen, K. L., Time‐dependent morphology changes in thin silver films on mica: A scaling analysis of atomic force microscopy results. *The Journal of Chemical Physics* **1996,** *105* (13), 5542-5551.
47. Wackenhut, F.; Failla, A. V.; Meixner, A. J., Multicolor Microscopy and Spectroscopy Reveals the Physics of the One-Photon Luminescence in Gold Nanorods. *J. Phys. Chem. C* **2013,** *117* (34), 17870-17877.

# Supporting information for:
# Hypericin: Single molecule spectroscopy of an active natural ingredient

Quan Liu,[1,3] Frank Wackenhut,[1*] Otto Hauler,[2] Miriam Scholz,[2] Pierre-Michel Adam,[3] Marc Brecht[2] and Alfred J. Meixner[1*]

[1] Institute of Physical and Theoretical Chemistry, Eberhard Karls University Tübingen, Germany

[2] Reutlingen Research Institute, Process Analysis and Technology (PA&T), Reutlingen University, Germany

[3] Laboratoire de Nanotechnologie et d'Instrumentation Optique, Université de Technologie de Troyes, France

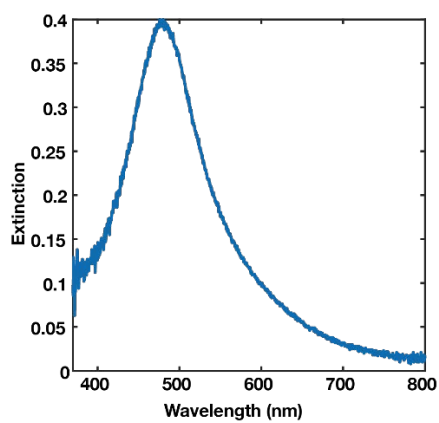

Figure S1: Extinction spectrum of the SERS substrate with a particle plasmon band at 482 nm.

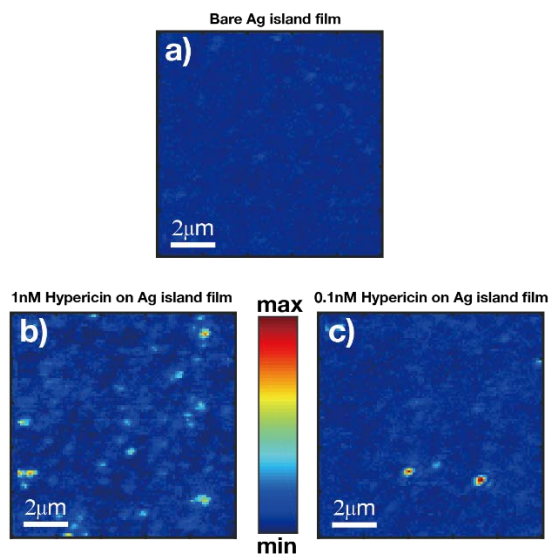

Figure S2: SERS scanning image. (a) shows scanning image of Ag island film before deposition of hypericin; (b,c) show SERS signal images after drop-casting of 1nM and 0.1nM Hypericin on Ag island film, respectively. After deposition of hypericin diffraction limited and intense hot spots can be observed, which are used to acquire the single molecule SERS spectra shown in Figure 5 (c) and (d). The excitation laser was 530nm (0.4 µW), scanning range was 10µmX10µm for all images.

Table 1: SERS peak positions in cm$^{-1}$ and relative intensity (in brackets) is given for the ensemble spectrum shown in Fig. 5(b), the deprotonated (Hyp⁻) and neutral (Hyp) SM shown in Fig. 5(c) and the respective simulated spectra.

| Ensemble | Exp. Hyp⁻ | Sim. Hyp⁻ | Exp. Hyp | Sim. Hyp | assignment[8] |
|---|---|---|---|---|---|
| 309 (0.04) | | 319 (0.28) | | 321 (0.09) | |
| 358 (0.04) | | 342 (0.35) | | 349 (0.20) | |
| 450 (0.35) | 448 (0.19) | 438 (0.39) | 455 (0.19) | 459 (1.00) | Skeletal deformation |
| 476 (0.17) | | 465 (0.37) | | 490 (0.13) | Skeletal deformation |
| | 529 (0.18) | 518 (0.11) | 518 (0.10) | 510 (0.13) | Skeletal deformation |
| | | | 593 (0.08) | 600 (0.08) | Skeletal deformation |
| 630 (0.13) | 636 (0.32) | 639 (0.12) | 637 (0.15) | 636 (0.05) | |
| 665 (0.09) | | 667 (0.08) | 746 (0.09) | | |
| 699 (0.07) | | 682 (0.08) | | 700 (0.04) | |
| 763 (0.11) | | | | | |
| 816 (0.07) | 830 (0.47) | 808 (0.04) | 790 (0.14) | 814 (0.02) | |
| 862 (0.07) | 868 (0.56) | 848 (0.08) | 839 (0.14) | 850 (0.09) | |
| 933 (0.12) | 938 (0.13) | 918 (0.17) | 939 (0.13) | 936 (0.17) | |
| 1018 (0.03) | 1017 (0.10) | 1017 (0.02) | 1016 (0.12) | 1025 (0.02) | |
| 1134 (0.09) | 1129 (0.67) | 1122 (0.17) | 1125 (0.44) | 1119 (0.16) | C-H bending |
| 1189 (0.13) | 1168 (0.56) | 1177 (0.16) | | 1190 (0.12) | C-O stretching |
| 1251 (0.44) | 1232 (0.70) | 1223 (0.24) | 1230 (0.63) | 1225 (0.24) | Ring in plane |
| 1297 (1.00) | 1299 (1.00) | 1298 (1.00) | | 1298 (0.46) | Ring in plane |
| 1333 (0.77) | 1364 (0.82) | 1324 (0.69) | 1316 (0.76) | 1327 (0.97) | Ring in plane |
| 1380 (0.66) | 1405 (0.46) | 1380 (0.26) | 1398 (0.92) | 1395 (0.22) | Ring in plane |
| 1450 (0.24) | 1471 (0.33) | 1467 (0.20) | 1451 (0.39) | 1444 (0.20) | Ring in plane |
| 1514 (0.19) | 1525 (0.24) | 1514 (0.17) | 1480 (0.28) | 1514 (0.40) | |
| | 1589 (0.87) | 1564 (0.21) | 1545 (0.57) | 1571 (0.58) | Ring stretching with C=O |
| 1604 (0.30) | 1613 (0.49) | 1614 (0.07) | 1610 (1.00) | 1609 (0.05) | C=O stretching |
| 1648 (0.29) | | 1625 (0.07) | | 1633 (0.13) | |